\newcommand{\R}{{\mathbb R}}   

\def\T{\tiny\mbox{\rm T}}

\newcommand{\vectf}[4]{\left[\begin{array}{c} #1 \\  #2 \\  #3 \\ #4 \end{array}\right]}

\newcommand{\mthree}[9]{\setlength\arraycolsep{5pt}
                        \left[\begin{array}{ccc}#1 & #2 & #3 \\
                                                #4 & #5 & #6 \\
                                                #7 & #8 & #9\end{array} \right]}

\def\bfa{\bm a}
\def\bfb{\bm b}
\def\bfc{\bm c}

\def\bfq{\bm q}

\def\bfu{\bm u}
\def\bfv{\bm v}
\def\bfw{\bm w}
\def\bfx{\bm x}
\def\bfy{\bm y}

\def\bfP{\bm P}
\def\bfQ{\bm Q}
\def\bfR{\bm R}
\def\bfS{\bm S}

\def\bfU{\bm U}
\def\bfV{\bm V}
\def\bfW{\bm W}
\def\bfX{\bm X}

\def\bfSigma{\bm\varSigma}

\def\bftheta{\bm\theta}

\documentclass[11pt, letterpaper]{article}

\usepackage{graphicx}
\usepackage{amsmath}
\usepackage{amssymb}
\usepackage{setspace}
\usepackage{epstopdf}
\usepackage{color}
\usepackage[letterpaper,left=1in,right=1in,top=1in,bottom=1in]{geometry}
\usepackage[linesnumbered,ruled,vlined,longend]{algorithm2e}
\usepackage{multirow}
\usepackage{longtable}
\usepackage{rotating}
\usepackage{threeparttable}
\usepackage{colortbl}
\usepackage{amsthm}
\usepackage[modulo]{lineno}
\usepackage{enumerate}
\usepackage{bm}

\newtheorem{remark}{Remark}

\title{\Large \bf A sensitivity-based approach to optimal sensor selection for process networks}

\author{\centerline{\normalsize Siyu Liu$^{a,b}$, Xunyuan Yin$^c$, Zhichao Pan$^{a}$, Jinfeng Liu$^{b,}$\thanks{Corresponding author: J. Liu. Tel: +1-780-492-1317. Fax: +1-780-492-2881. Email: jinfeng@ualberta.ca}}\vspace{5mm}\\ 
    \centerline{\small $^{a}$ School of Internet of Things Engineering, Jiangnan University, Wuxi\ 214122, China}\\
    \centerline{\small $^{b}$ Department of Chemical \& Materials Engineering, University of Alberta,}\\
    \centerline{\small Edmonton, AB, Canada, T6G 1H9}\\
    \centerline{\small $^{c}$ School of Chemical and Biomedical Engineering, Nanyang Technological University,}\\
    \centerline{\small 62 Nanyang Drive, Singapore, 637459}}
\allowdisplaybreaks

\begin{document}

\date{}

\maketitle
\setstretch{1.39}

\begin{abstract}

Sensor selection is critical for state estimation, control and monitoring of nonlinear processes. However, evaluating the performance of each possible combination of $m$ out of $n$ sensors is impractical unless $m$ and $n$ are small. In this paper, we propose a sensitivity-based approach to determine the minimum number of sensors and their optimal locations for state estimation. The local sensitivity matrix of the measured outputs to initial states is used as a measure of the observability. The minimum number of sensors is determined in a way such that the local sensitivity matrix is full column rank. The subset of sensors that satisfies the full-rank condition and provides the maximum degree of observability is considered as the optimal sensor placement. Successive orthogonalization of the sensitivity matrix is conducted in the proposed approach to significantly reduce the computational complexity in selecting the sensors. To validate the effectiveness of the proposed method, it is applied to two processes including a chemical process consisting of four continuous stirred-tank reactors and a wastewater treatment plant. In both cases, the proposed approach can obtain the optimal sensor subsets.
\end{abstract}


\section{Introduction} \label{sec:introduction}

The sensor selection problem arises in various applications, including target tracking \cite{Shen2014_IEEETSP,Anvaripour2019_TII}, agro-hydrological systems \cite{Nahar2019_CCE,Sahoo2019_AIChE}, power systems \cite{Qi2015_IEEETPS,Abooshahab2021} and wireless networks \cite{Mo2011_Auto,Bajovic2011_TSP}. Some useful methods on sensor selection have been developed. One straightforward solution to sensor selection is to perform an exhaustive search, which evaluates the performance given by each possible combination of $m$ out of $n$ candidate sensors \cite{Contreras2016_IEEETITS}. The number of combinations for an exhaustive search is $\frac{n!}{m!(n-m)!}$. Therefore, this type of solution is not suitable for large-scale systems. In \cite{Yin2018_CCE}, sensor selection is performed for the reduced-order WWTP model with small $m$ and $n$, in which case the exhaustive search is feasible. If sensors need to be selected from a large number of candidate sensors, then it may be difficult or even impossible to find the optimal solution.

Graph-based methods are widely used for the sensor node selection in networked systems based on the structural observability theory \cite{Stigter2017_SR}. By examining the strongly connected components (SCCs) in a graph describing node connections, graph-based methods can determine the number and locations of sensor nodes for ensuring structural observability \cite{Liu2013_PANS,Angulo2020_IEEETNSE}. The minimal set includes one sensor in each root SCC (an SCC with no incoming edges) of the observability inference diagram. Graph-based methods has become an influential method and has been used to provide insights into the relation between network topology and observability. However, since these approaches do not explicitly take into account model parameters, in some cases, the estimator designed based on the sensors selected by the graph-based approaches may lead to less accurate state estimates \cite{cowan2012nodal,Haber2018_IEEETCNS}. A workaround is to simultaneously deal with the sensor selection and state estimation, which has the potential to improve the overall estimation performance \cite{Haber2021_IEEETNSE,Nugroho2021_Auto}.

Another line of research on sensor selection is to rely on suitable observability criteria that are based on the properties of the empirical observability Gramian, e.g., the trace \cite{Singh2006_IECR}, the determinant \cite{Qi2015_IEEETPS}, and the smallest eigenvalue \cite{Kang2012}. The trace of an empirical observability Gramian matrix can be used as a measure of observability. However, a system can be unobservable if it has an eigenvalue of zero, despite a potentially large trace. The direction that has the lowest degree of observability can be determined by the smallest eigenvalue and the corresponding eigenvector. When the determinant of a Gramian matrix is used as a measure of observability, the smallest eigenvalue can be used as an additional measure in order to avoid the situation that the eigenvalue of Gramian matrix is almost zero and the system is not fully observable. A main limitation of the above methods is that the computation will become very expensive when applied to nonlinear systems, because these methods would require simulating the system many times for each perturbed initial condition in order to compute the ensemble average.

The observability of a nonlinear system can be determined by using the Lie derivative \cite{Mar2003}. However, this type of method only determines whether a system is observable but not the degree of observability. Also, due to the heavy computational burden, the Lie derivative is only suitable for small-scale systems. Sensitivity analysis can used as a tool for assessing the degree of observability, which has been often applied to quantitatively investigate parameter effects \cite{Kravaris2013_CCE,Miao2011_SIAM,Stigter2019_Auto} and simultaneous state and parameter estimation \cite{Liu2021_IECR,LiuSY2022_ChERD} but yet to be used for sensor selection.

In this work, we propose a sensitivity-based approach to select a minimum sensor set in a computationally efficient manner. A local sensitivity matrix describing the sensitivities between states and output measurements is constructed, which reflects the degree of observability of the different states. Once the minimum number of sensors to make the local sensitivity matrix full rank is determined, we perform successive orthogonalization of each column of the sensitivity matrix to rank sensors, and propose a measure of the degree of observability to determine the optimal sensor subset that satisfies the full-rank condition and maximizes the degree of observability. The proposed sensitivity-based approach avoids solving mixed-integer programming and exhaustive search, making it highly computationally efficient. Compared the aforementioned existing methods, the proposed approach maximizes the system observability and better correlates with the sensitivities that guide the development of the degree of observability.

The remainder of this paper is organized as follows. Section \ref{Section 2} describes the sensor selection problem and introduces two types of the existing methods. The construction of sensitivity matrix is shown in Section \ref{Section 3}. Section \ref{Section 4} defines the criteria for degree of observability and performs optimal sensor selection. In Section \ref{Examples}, we show the effectiveness of the proposed approach based on a chemical process consisting of four continuous stirred-tank reactors, and a wastewater treatment plant. The conclusions are drawn in Section \ref{Conclusions}.


\section{Preliminaries}
\label{Section 2}

\subsection{System description and problem formulation}
Consider the following nonlinear systems
\begin{equation}
\label{lsy10_2.1a}
 \bfx(k+1)=f(\bfx(k),\bfu(k))+\bfw(k),
\end{equation}
where $\bfx(k)\in\R^n$ denotes the state vector, $\bfu(k)\in\R^r$ is the input vector, $\bfw(k)\in\R^n$ denotes the system disturbance, and $f$ represents the state transition function.

The first objective of sensor selection is to determine the minimum number of sensors such that the system is observable; the second objective is to determine the optimal locations of the sensors. Consider a set of sensors $\mathcal{S}^{(m)}$ consisting of $m$ sensors such that the output $y_i(k)\in\R$ of sensor $i\in\mathcal{S}^{(m)}$ is given by
\begin{equation}
\label{lsy10_2.1b}
 y_i(k)=h_i(\bfx(k))+v_i(k),
\end{equation}
where $h_i(\bfx(k))\in\R$ is the sensor output function corresponding to sensor $i$, $v_i(k)\in\R$ denotes measurement noise. Define the output vector  $\bfy(k):=[y_1(k), y_2(k), \ldots, y_m(k)]^{\T}\in\R^m$ and noise vector $\bfv(k):=[v_1(k), v_2(k), \ldots, v_m(k)]^{\T}\in\R^m$. Then, we formulate the following measurement equation containing the measurements of the $m$ sensors
\begin{equation}
\label{lsy10_2.1c}
 \bfy(k)=h(\bfx(k))+\bfv(k).
\end{equation}
In the following, we seek to select an optimal sensor subset $\mathcal{S}^{(o)}\subseteq\mathcal{S}^{(m)}$, $o\leq m$ from the potentially available measured outputs in Equation (\ref{lsy10_2.1c}). For solving the sensor selection problem, some existing methods are summarized in the next subsection.

\subsection{Existing approaches}

Sensor selection problems on linear systems have attracted extensive efforts over the years. More recently, methods to address the nonlinear systems have also emerged. We will first briefly review the works on linear systems, followed by a more thorough review of the efforts on nonlinear systems.

For linear systems, efforts have been made to address the problem of sensor selection by minimizing the error in estimating its states. Such selection problems are in general NP-hard. For example, Zhang et al., minimized the trace of a priori and a posteriori error covariance matrices produced by a Kalman filter, and they showed that the optimization problem is NP-hard \cite{Zhang2017_Auto}. Another widely used method is greedy algorithms, a class of approximation algorithms. Such methods are guaranteed to have the desirable performance if the cost function is submodular \cite{Jawaid2015_Auto}.

For nonlinear networks, the sensor placement is typically formulated as a mixed-integer programming problem, for example, to maximize the determinant of the empirical observability Gramian as follows \cite{Qi2015_IEEETPS}:
\begin{subequations}
\label{lsy10_2.1d}
\begin{align}
\label{lsy10_2.1d1}
 &\max_{\bftheta}\det \bfW(\bftheta),\\
 &{\rm s.t.} \sum_{i=1}^{m}\theta_i=\bar{m}, \\
 &\theta_i\in\{0, 1\},\quad i=1,2,\ldots,m,
\end{align}
\end{subequations}
where $\bfW(\bftheta)$ denotes the empirical observability Gramian which is dependent on the value of the parameter $\bftheta$, $\bftheta\in\{0, 1\}^m$ is a binary decision vector to denote which sensors are selected, i.e., $\theta_i=1$ if $i\in\mathcal{S}^{(m)}$ is selected and $\theta_i=0$ otherwise, $\bar{m}$ is the number of sensors to be placed. The empirical observability Gramian $\bfW(\bftheta)$ is defined using the original system model, which reflects the observability of the system state in a given domain. However, the cost of computing the empirical observability Gramian is high because it requires simulating the process networks for different perturbations of the initial condition.

In (\ref{lsy10_2.1d1}), the determinant of the observability Gramian is considered as a measure of the degree of observability. In addition to the determinant, other measures based on the empirical observability Gramian are also used including the trace \cite{Singh2006_IECR}, the smallest eigenvalue \cite{Kang2012}, and the condition number \cite{Contreras2016_IEEETITS} of the Gramian matrix. A comparison of these degree of observability measures is reported in \cite{Qi2016}.

There are some other variants for the optimization problem in (\ref{lsy10_2.1d1}), for example, using the Jacobian matrix to replace the empirical observability Gramian matrix. Nonetheless, these variations are still mixed-integer programming problem \cite{Haber2018_IEEETCNS}. mixed-integer nonlinear programming problems are in general challenging to solve especially when the number of potential sensors is large. The solution proposed in this paper is operative and less computationally demanding. In the following section, we will elaborate on the method for selecting the optimal sensor set.

\section{Output to initial state sensitivity and observability test}
\label{Section 3}


In this work, we propose to use the sensitivity matrix of the output $\bfy$ to the initial state $\bfx(0)$ as a measure of the observability and introduce a computationally efficient optimal sensor selection algorithm based on the sensitivity matrix. In this section, we describe how the sensitivity matrix can be constructed and the corresponding observability test.

\subsection{Construction of the sensitivity matrix}

In sensor selection, we need to determine a sensor set such that the state of the system is observable (or can be constructed) based on the output information provided by the selected sensors. For a nonlinear system, the local observability of the state (or the initial state $\bfx(0)$) can be checked using the sensitivity matrix of the outputs to the initial state \cite{Grubben2018_IJC}. To construct the sensitivity matrix, we need the sensitivity of each output to each of the state element of $\bfx(0)$ as follows:
\begin{align}
\label{lsy10_3.1a}
	s_{ij}(k)=\frac{\partial y_i(k)}{\partial x_j(0)}=\frac{\partial h_i}{\partial x_j}(k)\bfS_{x,x(0)}(k),
\end{align}
where $\bfS_{x,x(0)}(k):=\frac{\partial \bfx(k)}{\partial \bfx(0)}$ is the sensitivity of the state $\bfx(k)$ with respect to the initial condition $\bfx(0)$. The sensitivity coefficient describes how sensitive the system output is to changes in a single state. At sampling time $k$, a time dependent matrix $\bfS_{y,x(0)}(k):=\frac{\partial \bfy(k)}{\partial \bfx(0)}$ is constituted where each column denotes the sensitivity of different outputs to each state element:
\begin{align}
\label{lsy10_3.1b}
	\bfS_{y,x(0)}(k)=&\mthree{s_{11}(k)}{\cdots}{s_{1n}(k)}
	{\vdots}{\ddots}{\vdots}
	{s_{m1}(k)}{\cdots}{s_{mn}(k)}\in\R^{m \times n}.
\end{align}
The above sensitivity $\bfS_{y,x(0)}(k)$ provides the sensitivity information based on the measurements obtained at time $k$. By collecting the sensitivity of all the past measurements from sampling time $0$ to sampling time $k$, the following sensitivity matrix can be constructed:
\begin{align}
\label{lsy10_3.1c}
	\bfS(k,0)=\vectf{\bfS_{y,x(0)}(0)}{\bfS_{y,x(0)}(1)}{\vdots}{\bfS_{y,x(0)}(k)}\in\R^{(k+1)m \times n},
\end{align}
The sensitivity matrix should be normalized before using it to check the observability so that the rank calculation will not be affected by the parameter values. Each column of the normalized sensitivity $\bfS(k,0)$ is a list of output sensitivities over time to the same element in the state vector, indicating the direction of sensitivity for that state element. If the norm of a column is larger than other columns, the output $\bfy$ is most sensitive to changes of the state element corresponding to the column.

\subsection{Observability test}
\label{Obtest}

This local sensitivity matrix is similar to the observability matrix of a linear system and can be used to check the local observability of a nonlinear system \cite{Liu2021_IECR}. Therefore, the local observability test can be formulated based on the rank test of the sensitivity matrix $\bfS(k,0)$. A full-rank sensitivity matrix indicates local observability. Singular value decomposition (SVD) is used for evaluating the rank of a given matrix. Specifically, SVD is performed for $\bfS(k,0)$ as follows:
\begin{align}
 \bfS(k,0)=\bfU \bfSigma \bfV^{\T},
\end{align}
where $ \bfU\in\R^{(km+m)\times (km+m)} $ and $ \bfV\in\R^{n\times n} $ are orthonormal matrices, $ \bfSigma\in\R^{(km+m)\times n} $ contains the $n$ singular values $\{\sigma_r, r=1,2,\dots,n\}$ of $\bfS(k,0)$ in decreasing order on the diagonal. If $\sigma_n=0$, or the smallest singular value is smaller than a pre-determined threshold, $\bfS(k,0)$ is rank-deficient.

In the proposed sensor selection algorithm, observability test is conducted to find the sensor combinations that can make the system state observable. In addition to observability, we also resort to the degree of observability to determine the optimal sensor selection in the next section.

\section{Proposed optimal sensor selection algorithm}
\label{Section 4}

Based on the local sensitivity matrix, we propose a measure to quantify the degree of observability of different combinations of sensors based on successive orthogonalization. Then, sensors are ranked based on their contributions to the overall degree of observability. The minimum sensor set and optimal locations of sensors are determined through maximizing the degree of observability of the system while ensuring the local sensitivity matrix is full column rank.

\subsection{A measure for degree of observability}

Since each column of the sensitivity matrix contains the sensitivities of the output $\bfy$ to a state element, it is proposed to use the norm of each column for the initial screening of the degree of observability of each of the state elements. The output is most sensitive to the state element that corresponds to the column with the maximum norm. This also implies that the output measurements carry the most information of this state element and this state element is most observable based on the output measurements.

While the norm of the columns is useful in identifying the most observable state element, it should not be used directly to examine the other columns due to the potential linear dependence of the columns. The highly linear correlation between any two columns suggests that changes in the corresponding state elements would have similar effects on the outputs. Thus, the two state elements are indistinguishable based on the measurements. It is proposed to use the orthogonalization method to check the dependence of the columns of the sensitivity matrix. The idea of this method is to calculate the perpendicular distance of one column to the vector space spanned by the other columns and then uses the distance as a measure of the linear dependency. By successive orthogonalization, it is possible to rank the columns in terms of their information contained in the measurements. Also, the proposed measure for degree of observability is based on total information contained in the measurements regarding the state elements.

\begin{figure}[!hbt]
 \centering
 \includegraphics[width=0.7\hsize]{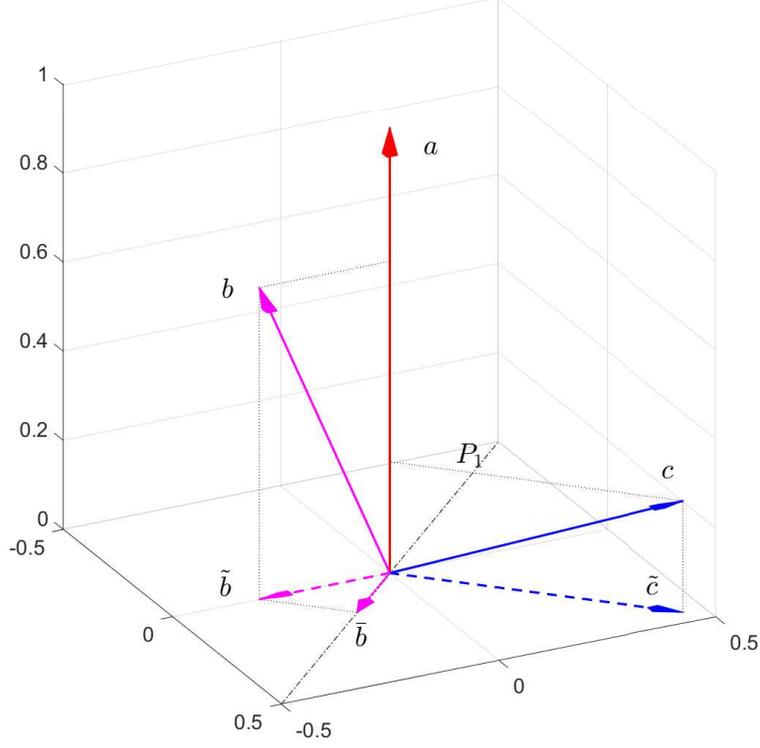}
 \caption{Illustration of the orthogonolization }
\label{lsy10_fig_ortho}
\end{figure}
The way to calculate the proposed measure is illustrated in the following example with three elements in the state vector. Consider that the sensitivity matrix $\bfS(k,0):=[\bfa, \bfb, \bfc]$ where $\bfa$, $\bfb$, $\bfc$ are the columns. In the first step, we calculate the norm of each of the three column vectors. Without loss of generality, we assume the vector $\bfa$ has the largest norm denoted as $F_1$. $F_1$ is an indicator of the amount of information contained in the measurements regarding the state element corresponding to $\bfa$. Due to potential linear dependence between the columns, the other two columns $\bfb$, $\bfc$ may also contain information that is
already expressed by column $\bfa$. To proceed, we define $\bfX_1:=\bfa$. The direction of $\bfX_1$, $\bfq_1=\bfX_1/\|\bfX_1\|$ is used as the first basis for the orthogonalization.

In the second step, the other two vectors are projected onto $\bfq_1$. The information in $\bfS(k,0)$ that cannot be represented by $\bfX_1$ can be calculated by
\begin{align}
\tilde{\bfb}=\bfb-(\bfq_1^{\T}\bfb)\bfq_1,\\
\tilde{\bfc}=\bfc-(\bfq_1^{\T}\bfc)\bfq_1.
\end{align}
$\tilde{\bfb}$ and $\tilde{\bfc}$ are now orthogonal to $\bfq_1$. Let us define the residual matrix $\bfR_1:=[\tilde{\bfb},\tilde{\bfc}]=\bfS(k,0)$ $-\bfX_1(\bfX_1^{\T}\bfX_1)^{-1}\bfX_1^{\T}\bfS(k,0)$, which represents the information that cannot be captured by the first basis in the sensitivity matrix.

In the third step, we evaluate the norm of each column of $\bfR_1$. Suppose that $\tilde{\bfc}$ has a larger norm $F_2$. $F_2$ indicates the extra information contained in the measurements regarding the state element corresponding to $\bfb$ that has not been covered by column $\bfa$. Next, we add $\bfc$ from $\bfS(k,0)$ to $\bfX_1$ to form a new matrix with two columns $\bfX_2=[\bfa,\bfc]$. The second basis $\bfq_2=\tilde{\bfc}/\|\tilde{\bfc}\|$ is formed and the information that cannot be expressed by $\bfX_2$ is evaluated by projecting $\tilde{\bfb}$ to the line/plane (denoted $P_1$) that is perpendicular to $\bfX_2$. The new vector
\begin{align}
	\bar{\bfb}=\tilde{\bfb}-(\bfq_2^{\T}\tilde{\bfb})\bfq_2
\end{align}
with the minimum norm $F_3$, which represents the extra information contained in the measurements regarding the state element corresponding to $\bfb$ that is not captured by $\bfa$ or $\bfc$. The corresponding residual matrix is $\bfR_2:=[\bar{\bfb}]=\bfS(k,0)-$ $\bfX_2(\bfX_2^{\T}\bfX_2)^{-1}\bfX_2^{\T}\bfS(k,0)$. By appending $\bfb$ to $\bfX_2$, we can form a new matrix $\bfX_3=[\bfa,\bfc,\bfb]$. In $\bfX_3$, the columns are ordered according to their contributions to the information contained in the measurements regarding the state elements. The above procedure is also illustrated in Figure~\ref{lsy10_fig_ortho}.

For a sensor set $\mathcal{S}^{(o)}$, we assume that $\mathcal{S}^{(o)}=\mathcal{S}^{(m)}=\{1,2,\dots,m\}$, and we can construct the sensitivity matrix $\bfS(k,0)\in\R^{(km+m)\times n}$ according to the sensor measurements $\bfy$ provided by $\mathcal{S}^{(o)}$. The projection is repeated for $m-1$ steps, and we can obtain multiple norms denoted as $F_i$, $i=1,2,\dots, m$. These $F_i$ are indicators of the amount of information contained in the measurements regarding the state elements. Let us define
\begin{align}
	\label{lsy10_4.2a}
	F=\sum_{j=1}^{m}F_j.
\end{align}
In (\ref{lsy10_4.2a}), $F$ represents the amount of information contained in the measurements provided by the current sensor combination $\mathcal{S}$ regarding the state. We propose to use $F$ as the measure of the degree of observability of the state. However, at this time, whether the system is observable is not considered. According to the SVD presented in Subsection \ref{Obtest}, we calculate the rank of the current sensitivity matrix $\bfS(k,0)$. The proposed measure for degree of observability is as follows:
\begin{align}
\label{lsy10_3.2a}
 D(\mathcal{S}^{(o)})=\alpha F,
\end{align}
where
\begin{equation}
	\label{eq6}
	\alpha = \left\{
	\begin{aligned}
		1 & , & {\rm rank}(\bfS(k,0)) = n, \\
		0 & , & {\rm rank}(\bfS(k,0)) < n.
	\end{aligned}
	\right.
\end{equation}
If rank($\bfS(k,0))=n$, then $\alpha=1$. The degree of observability $D(\mathcal{S}^{(o)})$ is effective; If rank($\bfS(k,0))\textless n$, $\alpha$ is set to zero, then $D(\mathcal{S}^{(o)})$ is also equal to zero. This implies that we should only evaluate the degree of observability of observable systems.

The procedure of calculating the degree of observability using the orthogonalization method is summarized in Algorithm \ref{degree}:

\begin{algorithm} \label{degree}
	\caption{Calculating the degree of observability}
	\KwIn{ Sensor set $\mathcal{S}^{(o)}$, sensitivity matrix $\bfS(k,0)$ based on $\mathcal{S}^{(o)}$}
	\KwOut{Degree of observability $D(\mathcal{S}^{(o)})$ }
	
	Calculate the rank of normalized $\bfS(k,0)$.\\
	\If{{\rm rank}$(\bfS(k,0)) < n$}{
	$\alpha=0$}
	\Else{$\alpha=1$}
	Calculate the norm of each column of the normalized $\bfS(k,0)$.\\
	Select the column with the largest norm $F_1$, and denote it as $\bfX_1$.\\
    \For{$l=1:m-1$}
    {
    Calculate the residual matrix: $\bfR_l=\bfS(k,0)-\bfX_l(\bfX_l^{\T}\bfX_l)^{-1}\bfX_l^{\T}\bfS(k,0)$.\\
    Select the column from $\bfS(k,0)$ that corresponds to the column with the maximum norm $F_{l+1}$ in $\bfR_l$, and add it to $\bfX_l$ as a new column to form $\bfX_{l+1}$.}
    Calculate $F$ by (\ref{lsy10_4.2a}).\\
	Calculate $D(\mathcal{S}^{(o)})=\alpha F$. \\
	\textbf{return} $D(\mathcal{S}^{(o)})$
\end{algorithm}



\subsection{Selection of the minimum sensor set}

In this section, we discuss the proposed computationally efficient approach for determining the minimum number of sensors to ensure the observability of the state vector while maximizing the degree of observability given the number of the sensors.

In the proposed approach, we select the minimum sensor set by removing sensors from the available sensor set one by one based on the degree of observability. Initially, all the sensors ($\mathcal{S}^{(m)}$) are selected and the system is assumed to be observable. Then, the sensor that contributes the least to the degree of observability is determined and removed from the selected sensor set. To determine this specific sensor, we remove one sensor from the selected sensor set. There are $m$ possible options. For each of these $m$ options, we calculate the corresponding degree of observability. Among the $m$ degrees of observability, we find the maximum one. The sensor that is not included in the set corresponding to the maximum degree of observability is the sensor that contributes the least to the degree of observability and should be removed at this step. Following this way, we remove the sensors one by one until the system is not observability. At certain step, if removing a sensor makes the system unobservable, the algorithm stops. The last observable sensor set is the selected optimal minimum sensor set.

Below is a strategy for removing a single sensor at a time. The optimal sensor $i^*$ to be removed is selected by
\begin{equation}
\label{lsy10_4.2b}
i^*=arg\max_{i\in\mathcal{S}^{(o)}} D(\mathcal{S}^{(o)} \backslash \{i\}).
\end{equation}
The above strategy is to remove the sensor $i\in\mathcal{S}^{(o)}$ and calculate the degree of observability for each possible option. From $m$ options, the sensor to be removed is the solution of maximizing the degree of observability $D(\mathcal{S}^{(o)} \backslash \{i\})$. When the  sensor $i^*$ is removed, the current sensor set $\mathcal{S}^{(o)}=\mathcal{S}^{(o)}\backslash \{i^*\}$. To proceed, we will continue screening sensors from the current sensor $\mathcal{S}^{(o)}$ with $m-1$ sensors. At this time, the dimension of the sensitivity matrix will change with the updated sensor set $\mathcal{S}^{(o)}$. In the next step, there are $m-1$ possible options. For each of these $m-1$ options, we construct the sensitivity matrix $\bfS(k,0)\in\R^{(m-2)(k+1)\times n}$, and calculate the corresponding degree of observability. The sensor that is not included in the set corresponding to the maximum degree of observability is selected to remove from $m-1$ sensors at this step.

The procedure of the proposed method is summarized in Algorithm \ref{minimum_set} for solving two optimization objectives: 1) determining the minimum sensor set while ensuring the system observable; and 2) obtaining the maximum degree of observability of the minimum set.

\begin{algorithm} \label{minimum_set}
    \caption{Determine the best minimum sensor set}

    The set containing all sensors is $\mathcal{S}^{(m)}=\{1,2,\ldots,m\}$. The initial set is: $\mathcal{S}^{(o)}=\mathcal{S}^{(m)}$. \\

    Set $o=m$. Perform the following steps: \\

    \For{$i\in\mathcal{S}^{(o)}$}{
        Remove the $i$th sensor from $\mathcal{S}^{(o)}$.\\
        Perform {\bf Algorithm 1} and obtain $i$ degrees of sensitivity $D(\mathcal{S}^{(o)})$.\\}
        \If{{\rm all} $D(\mathcal{S}^{(o)})=0$}{
          Terminate the procedure.
        }
        \Else{
    Choose $i^*$ by (\ref{lsy10_4.2b}) and remove the corresponding sensor from $\mathcal{S}^{(o)}$.\\
    $\mathcal{S}^{(o)}\leftarrow\mathcal{S}^{(o)}\backslash \{i^*\}$, $o\leftarrow o-1$, go to Step 3. }

    \textbf{return} $\mathcal{S}^{(o)}$
\end{algorithm}

In Algorithm 2, according to the criterion for quantifying the degree of observability of different sensor combinations, we determine to remove the sensor that contributes less to the observability of the system based on the the degrees of observability. Therefore, the optimal sensor set is selected so that the degree of observability is maximized. The minimum number of sensors are determined for the observability of the system and the optimal sensor subset is determined on maximizing the degree of observability of the system.



\begin{remark}
	The sensor selection algorithm proposed in this paper is a good option from an economic perspective since it requires the minimum number of sensors. However, the minimum sensor set may not be able to provide sufficiently large degree of observability. In order to have robust state estimation performance in the presence of significant process disturbances and measurement noise, additional sensors can be added to the minimum set according to the ordering of sensors given by the algorithm. A detailed illustration is given through the second application in Section~\ref{Examples}.
\end{remark}

\begin{remark}
	The cost of different sensors is an important factor that may affect the result of sensor selection. Taking into account the cost will make the proposed approach more applicable, and it will be investigated in the future work.
\end{remark}

\subsection{Complexity of the algorithm}

It is well known that the main challenge associated with the sensor selection is the computational complexity. A comparison on the problem of computational complexity of some classical algorithms for the sensor selection, like global searching method, solving the mixed-integer programming, and greedy algorithm will be discussed briefly in the following. First of all, the procedure proposed in this work is computationally efficient by removing the sensors one by one from the sensor set based on the degree of observability. The number of all possible sensor combinations that needs to be considered is $m+(m-1)+\cdots+(o+1)+o=\frac{1}{2}(m-o+1)(m+o)$ (when selecting $o$ sensors from a set of $m$ sensors).
\begin{itemize}
\item As the global searching method in \cite{Yin2018_CCE}, when selecting minimum number sensors from a set that can make the system observable, the number of the sensors should be increased form $1$ to $m$ until the minimum number $o$ is found. Therefore, a total of $C_m^1+C_m^2+\cdots+C_m^o$ sensor combinations need to be checked.

\item The sensor selection is typically formulated as a mixed-integer programming problem, as shown in Equation (\ref{lsy10_2.1d}). Solving the mixed-integer programming is challenging especially for nonlinear cases. Even for the mixed-integer linear programming, it is still NP-hard. In particular, the special case of $0-1$ integer linear programming, in which unknowns are binary, the number of all possible solutions is $2^m$.

\item
The greedy method is another widely used algorithm for the sensor selection problem. For example, Zhang et al., applied the greedy methods to select sensors that yield the largest incremental reduction in steady-state (a priori or a posteriori) error covariance \cite{Zhang2017_Auto}. Similarly, an alternative strategy is to add the sensors one by one using the greedy method to maximize the degree of observability. This approach would have $m+(m-1)+\cdots+(m-o+1)=\frac{o}{2}(m-o+1)$ possible sensor combinations that needs to be considered.
\end{itemize}
Compared with above methods, when $m=2o-1$, the number of sensor combinations of the adding method is the same as the removing strategy, in which the computational complexity is acceptable. However, the adding strategy is severely flawed by the lack of meaning of observability in the early stages where there are too few sensors. That is, it is likely that the earlier sensor selection is not optimal and that a larger number of sensors must be selected. In contrast, the system in removal strategy is always observable by virtue of the method of eliminating the least contributing sensors. Thus the minimum number of sensors is guaranteed. For this consideration, the proposed removal strategy seems to be a superior method than the adding one.

\section{Simulation results}
\label{Examples}

In this section, we apply the proposed procedure and algorithms to determine the minimum number of sensors and the optimal sensor placement for two nonlinear process networks.

\subsection{Application to a four-CSTR process}

\begin{figure}[!hbt]
 \centering
 \includegraphics[width=0.8\hsize]{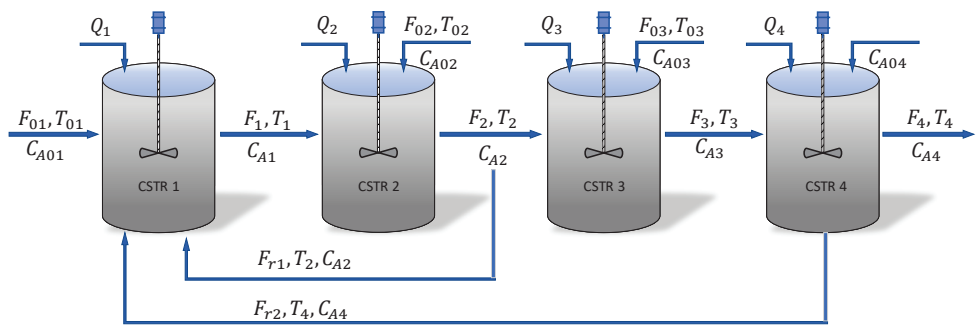}
 \caption{A schematic of the four-CSTR process}
\label{lsy10_fig_4CSTR}
\end{figure}

First, we consider a chemical process consisting of four continuous stirred-tank reactors (CSTRs), of which a schematic is shown in Figure~\ref{lsy10_fig_4CSTR}. Pure reactant $A$ is fed into the $i$th reactor for reactions at temperature $T_{0i}$, flow rate $F_{0i}$, and molar concentration $C_{0i}$, $i=1,2,3,4$. During the reaction process, the outlet flow of each of the first three reactors is supplied to its adjacent downstream tank. A portion of the outlet stream of the second reactor is recycled back to the first reactor at flow rate $F_{r1}$, temperature $T_2$ and molar concentration $C_{A2}$. The effluent of the fourth reactor is constituted by a recycle flow back to the first reactor and a discharging flow for further processing. Three parallel irreversible exothermal reactions $A\rightarrow B$, $A\rightarrow C$, and $A\rightarrow D$ take place in the reactors. The heat $Q_i$ is removed from/provided to the $i$th reactor by using the equipped jacket. Based on mass and energy balances, a first-principles process model has been established, which can be found in \cite{Rashedi2018_TII}.

The molar concentration $C_{Ai}$ and temperature $T_i$ in the $i$-th reactor are system states, $i=1,2,3,4$. Table~\ref{lsy10_taba1} shows the values of model parameters, where $V_i$, $E_i$, $\Delta H_1$, $k_i$ are the volume of reactors, the activation energies, enthalpies, and pre-exponential constants of the reactions, $\rho$ and $c_p$ denote the density and heat capacity of the mixture in the reactors.

\begin{table}[!hbt] \small
 \centering
 \caption{Parameter values of the four-CSTR process}
 \label{lsy10_taba1}
\renewcommand{\arraystretch}{1.2}
 \tabcolsep 14pt
 \begin{tabular}{l|l|l|l}\hline
  $T_{01}=300$ K   &  $C_{01}=4.0$ kmol/$\rm m^3$ & $k_1=3.0\times10^6$ $\rm h^{-1}$ & $\Delta H_1=-5.0\times 10^4$ kJ/kmol  \\
  $T_{03}=300$ K   &  $C_{02}=2.0$ kmol/$\rm m^3$ & $k_2=3.0\times10^5$ $\rm h^{-1}$ & $\Delta H_2=-5.2\times 10^4$ kJ/kmol  \\
  $T_{02}=300$ K   &  $C_{03}=3.0$ kmol/$\rm m^3$ & $k_3=3.0\times10^5$ $\rm h^{-1}$ & $\Delta H_3=-5.0\times 10^4$ kJ/kmol  \\
  $T_{04}=300$ K   &  $C_{04}=3.5$ kmol/$\rm m^3$ & $F_1=35$ $\rm m^3$/h    & $E_1=5.0\times10^4$ kJ/kmol \\
  $F_{01}=5$ $\rm m^3$/h   &  $V_1=1$ $\rm m^3$   & $F_2=45$ $\rm m^3$/h    & $E_2=7.5\times10^4$ kJ/kmol \\
  $F_{02}=10$ $\rm m^3$/h  &  $V_2=3$ $\rm m^3$   & $F_3=33$ $\rm m^3$/h    & $E_3=7.53\times10^4$ kJ/kmol\\
  $F_{03}=8$ $\rm m^3$/h   &  $V_3=4$ $\rm m^3$   & $F_{r1}=20$ $\rm m^3$/h & $R=8.314$ kJ/(kmol$\cdot$K)\\
  $F_{04}=12$ $\rm m^3$/h  &  $V_4=6$ $\rm m^3$   & $F_{r2}=10$ $\rm m^3$/h & $c_p=0.231$ kJ/(kg$\cdot$K)\\
  $\rho=1000$ kg/m$^3$     &  & &  \\\hline
 \end{tabular}\end{table}

In the simulations, heating inputs to the four reactors are chosen to be $Q_1=1.0\times10^4$kJ/h, $Q_2=2.0\times10^4$kJ/h, $Q_3=2.5\times10^4$kJ/h, and $Q_4=1.0\times10^4$kJ/h. The continuous model is discretized using the fourth-order Runge-Kutta method with a sample time $\Delta t=\frac{1}{120}$h. We assume that sensors are available to measure all the states ($m=N=8$). The objective is to find the optimal minimum sensor set. At each sampling point, the sensitivity matrix $\bfS(k,0)$ is calculated. Initially, the sensor set $\mathcal{S}^{(o)}=\mathcal{S}^{(m)}$ where $\mathcal{S}^{(m)}=\{C_{A1}, T_1, C_{A2}, T_2, C_{A3}, T_3, C_{A4}, T_4\}$ denoting as $\{1,2,3,4,5,6,7,8\}$ for simplicity, and the system is fully observable. Furthermore, we consider that the four-CSTR model is affected by the process noise with zero mean and standard deviation $0.1\bfx_s$ ($\bfx_s$ is the steady-state point of $\bfx$ corresponding to the constant values of the inputs picked earlier), and measurement noise with zero mean and standard deviation $0.1\bfy_s$ where $\bfy_s$ is the value of $\bfy$ at steady-state.


\begin{figure}[!hbt]
 \centering
 \includegraphics[width=0.8\hsize]{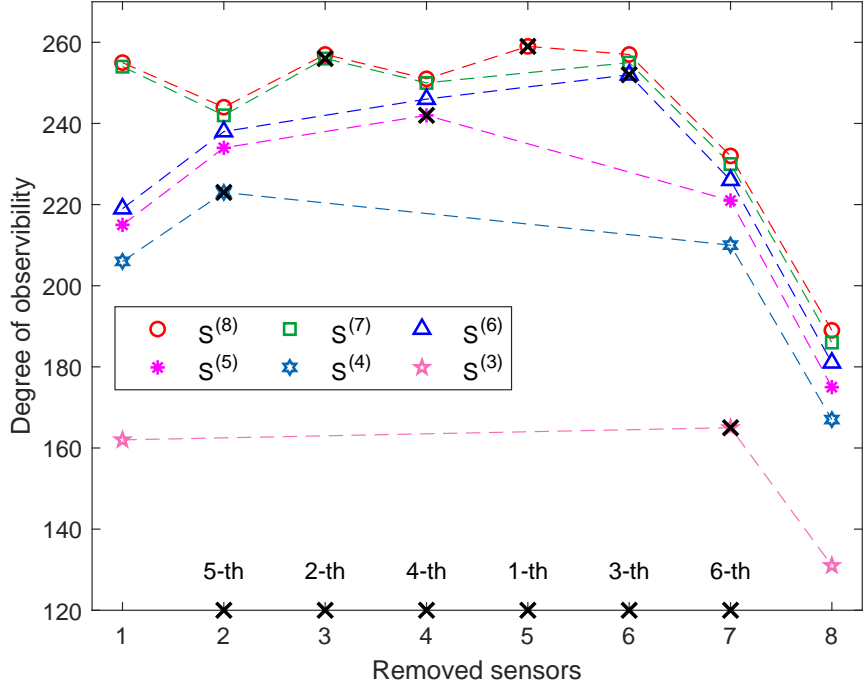}
 \caption{Degrees of observability by removing different sensor from remaining sensor set.}
\label{lsy9_figa1}
\end{figure}

\begin{table}[!hbt] \small
	\centering
	\caption{The process of sensor selection for four-CSTRs.}
	\label{lsy10_taba2}
	\renewcommand{\arraystretch}{1.2}
	\tabcolsep 5pt
	\begin{tabular}{cccccc}\hline
		$m$ & $\mathcal{U}$ & $\mathcal{S}^{(o)}$  & rank($\bfS(k,0)$) &  degree$_{\max}$ & sensor to remove  \\\hline
		8   & $\{\}$              & $\{1,2,3,4,5,6,7,8\}$&   8          &   261.47        &   5 \\
		7   & $\{5\}$             & $\{1,2,3,4,6,7,8\}$  &   8          &   259.61        &   3 \\
		6   & $\{5,3\}$           & $\{1,2,4,6,7,8\}$    &   8          &   255.97        &   6 \\
		5   & $\{5,3,6\}$         & $\{1,2,4,7,8\}$      &   8          &   251.65        &   4 \\
		4   & $\{5,3,6,4\}$       & $\{1,2,7,8\}$        &   8          &   241.89        &   2 \\
		3   & $\{5,3,6,4,2\}$     & $\{1,7,8\}$          &   8          &   223.15        &   7 \\
		2   & $\{5,3,6,4,2,7\}$   & $\{1,8\}$            &   8          &   164.74        &   1 \\
		1   & $\{5,3,6,4,2,7,1\}$ & $\{8\}$              &   5          &   67.7          &     \\\hline
\end{tabular}\end{table}
To find the optimal minimum sensor set, the sensors will be removed one by one from $\mathcal{S}^{(o)}$ by performing the proposed sensor selection algorithm. The corresponding degree of observability of remaining sensor combinations is shown in Figure~\ref{lsy9_figa1} and the steps are recorded in Table~\ref{lsy10_taba2} ($\mathcal{U}$ is the set of removed sensors, $m$ is the number of sensors in $\mathcal{S}^{(o)}$, the maximum degree of observability is denoted as degree$_{\max}$). The initial value of the degree of observability is 261.47 when all the sensors are selected. For $S^{(8)}$, as shown in Figure~\ref{lsy9_figa1}, when the fifth sensor is removed, the corresponding degree of observability of remaining seven sensors is the largest, which is 259.61 as shown in Table~\ref{lsy10_taba2}. This means that removing the fifth sensor has the smallest impact on the system observability. Therefore, the sensor for state 5 ($C_{A3}$) is removed in this step. In the next step, another sensor is to be removed and we aim to find the optimal set containing six sensors. It is found that if the third one is removed from $S^{(7)}$, the degree of observability is the largest (255.97) among all possible combinations of six sensors. Therefore, the third sensor is removed at this step. The remaining steps for removing sensors are similar until the 7th sensor is removed from $S^{(3)}$. From Table~\ref{lsy10_taba2}, it can be seen that degree$_{\max}$ value becomes smaller with the decrease of the number of sensors. In the case of ensuring the observability of the system, the decrease in degree$_{\max}$ value is not very significant (from $m=8$ to $m=2$). However, if we further remove any of the remaining two sensors, the rank is deficient (rank($\bfS(k,0)$)=5), and the degree$_{\max}$ value decreases dramatically to only 67.7. Therefore, the sensor selection process is terminated at $m=2$, and sensors $5,3,6,4,2,7$ is removed in sequence. It can be seen from Figure~\ref{lsy9_figa1} and Table~\ref{lsy10_taba2} that the selected sensor set is $\{1,8\}$, i.e., $\{C_{A1},T_4\}$, with the maximum degree of obserbability for the minimum number of sensors.


\begin{figure}[!hbt]
 \centering
 \includegraphics[width=0.8\hsize]{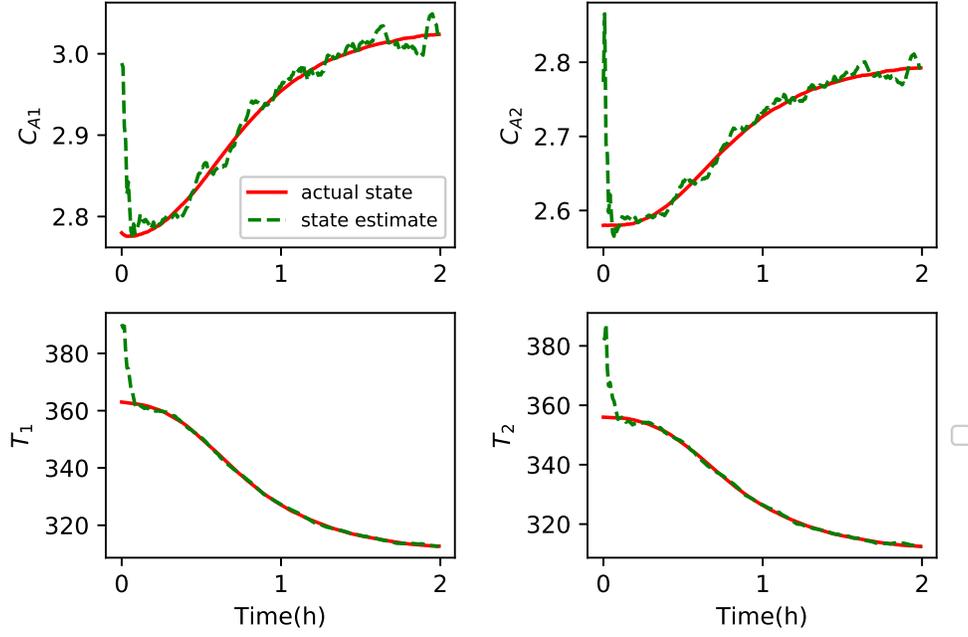}
 \caption{The trajectories of the actual states ($C_{A1}$, $T_1$, $C_{A2}$, $T_2$) and their state estimates based on the sensor set \{$C_{A1}$, $T_4$\}.}
\label{lsy9_figa2}
\end{figure}

\begin{figure}[!hbt]
	\centering
	\includegraphics[width=0.8\hsize]{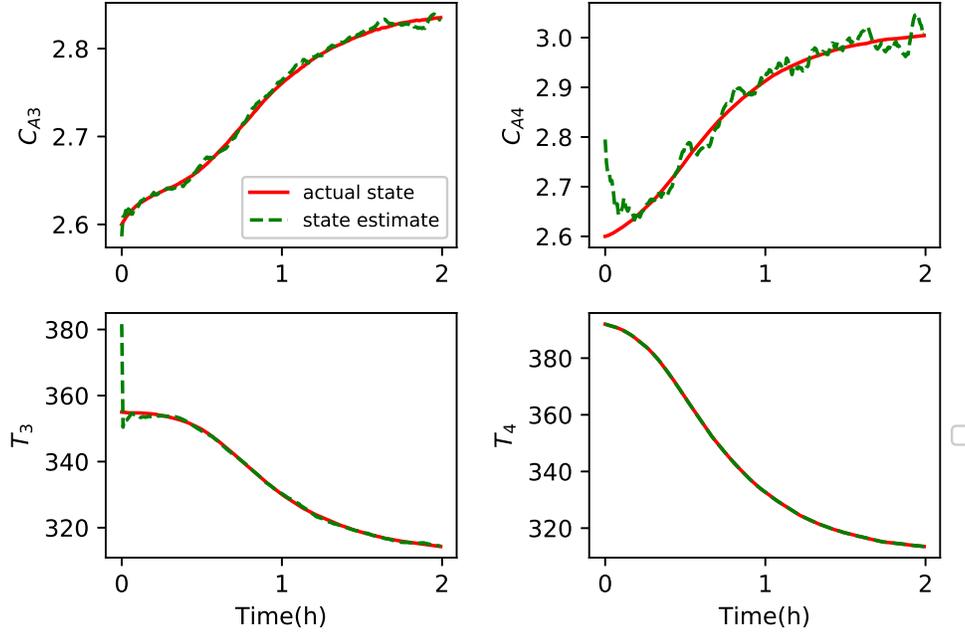}
	\caption{The trajectories of the actual states ($C_{A3}$, $T_3$, $C_{A4}$, $T_4$) and their state estimates based on the sensor set \{$C_{A1}$, $T_4$\}.}
	\label{lsy9_figa3}
\end{figure}

Now, we have determined the minimum number of sensors and the optimal placement of sensors. Next, we consider state estimation using the moving horizon estimation (MHE) to show that the selected sensor set $\{C_{A1},T_4\}$ provides good state estimation performance. The process and measurement noises are the same as that of the sensor selection algorithm. The input variables and sampling time are also the same as before. The output vector here is $ [C_{A1},T_4]^{\T} $ corresponding to the selected set. The steady-state value $\bfx_s:=[2.78, 363, 2.58, 356, 2.6, 355, 2.6, 392]^{\T}$, and the initial value guess is set to be $1.2\bfx_s$. The moving horizon length is 40. The covariance matrices of process noise and measurement noise in the MHE are $\bfQ_w={\rm diag}((0.1\bfx_s)^2)$, $\bfR_v={\rm diag}((0.1\bfy_s)^2)$. The actual states and the state estimates are shown in Figures~\ref{lsy9_figa2} and \ref{lsy9_figa3}. It can be observed that the state estimates are close to their actual states.

\begin{figure}[!hbt]
	\centering
	\includegraphics[width=0.8\hsize]{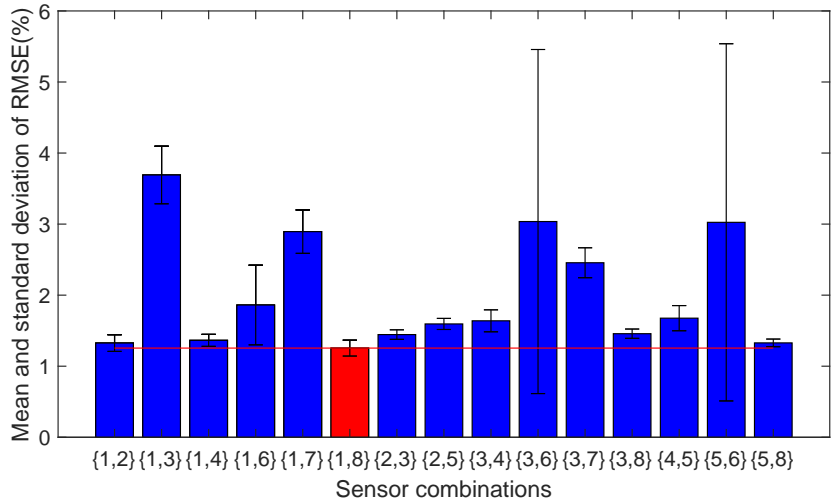}
	\caption{Mean and standard deviation of RMSEs with 10 runs for 15 sensor combinations with 1\% measurement and process noises }
	\label{lsy9_figa4}
\end{figure}

The root-mean-square error (RMSE) is used to evaluate the performance of state estimation. In order to reduce the impact of noise randomness on the estimation results, we can perform the state estimation several times and average it as an evaluation of performance. For further showing the validity of selected sensor set, all possible two sensor combinations are performed as a comparison. There are a total of $ C_8^2=28 $ sets, of which the mean and standard deviation of the RMSEs (based on 10 different simulation runs) for 15 sensor combinations are shown in Figure~\ref{lsy9_figa4}, and the corresponding degrees of the observability are shown in Table~\ref{lsy10_tab3}. The other 13 cases cannot make the system observable. In Figure~\ref{lsy9_figa4}, it can be found that the selected sensor set $\{1,8\}$ gives the smallest estimation errors compared to other sensor sets with the same number of sensors, and the degree of observability is largest in Table~\ref{lsy10_tab3}. For the sets $\{1,3\}$, $\{3,6\}$ and $\{5,6\}$, their mean values of RMSE in Figure~\ref{lsy9_figa4} are large, and the corresponding degrees of observability in Table~\ref{lsy10_tab3} are relatively small.
\begin{table}[!hbt] \small
	\centering
	\caption{Degree of observability}
	\label{lsy10_tab3}
	\renewcommand{\arraystretch}{1.2}
	\tabcolsep 8pt
	\begin{tabular}{lccccccccc}\hline
		& \{1,2\} & \{1,3\} & \{1,4\} & \{1,6\} & \{1,7\} & \{1,8\} & \{2,3\} & \{2,5\} \\\hline
		degree &  90.1   &   48.1  &   63.7  &  67.0   &   65.7  &  141.4  &   91.1  &   73.4  \\\hline
		& \{3,4\} & \{3,6\} & \{3,7\} & \{3,8\} & \{4,5\} & \{5,6\} & \{5,8\} \\\hline
		degree &  77.2   &   51.8  &   72.1  &  99.0   &   71.4  &   45.0  &   109.6 \\\hline
\end{tabular}\end{table}


\subsection{Application to a wastewater treatment plant}

\begin{figure}[!hbt]
	\centering
	\includegraphics[width=0.8\hsize]{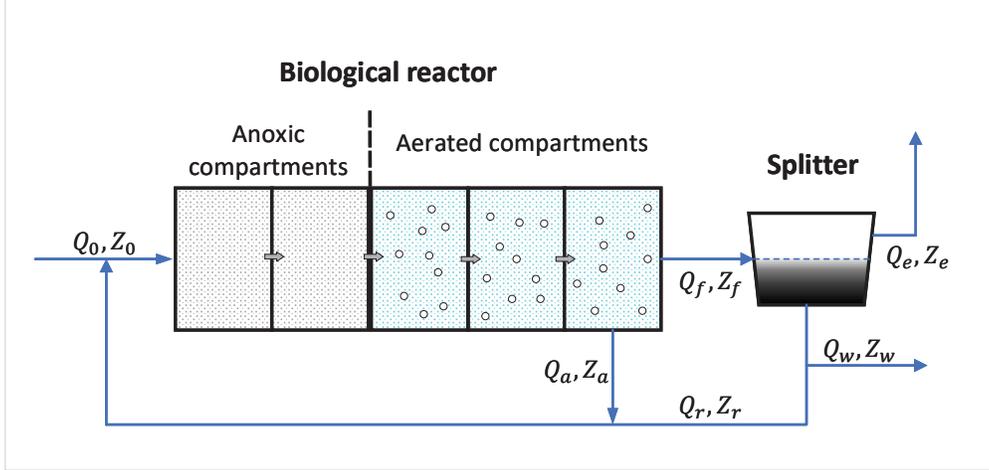}
	\caption{A schematic of the wastewater treatment plant }
	\label{lsy10_figb}
\end{figure}

In this subsection, we apply the proposed approach to a wastewater treatment plant (WWTP), which is a large-scale process network \cite{Alex2008}. A schematic diagram of the WWTP is presented in Figure~\ref{lsy10_figb}, which consists of a five-compartment activated sludge reactor and a spiltter. The biological reactor has two sections: the non-aerated section containing the first two anoxic chambers and the aerated section consisting of the remaining three chambers.

Wastewater is fed into the first anoxic reactor at concentration $Z_0$ and flow rate $Q_0$. A portion of the effluent with $Z_a$ and $Q_a$ of the last aerated reactor is recycled back to the first anoxic reactor and the rest of the effluent is fed into the splitter at concentration $Z_f$ and flow rate $Q_f$. The processed water leaves from the top of the splitter at concentration $Z_e$ and flow rate $Q_e$. The generated sludge is withdrawn from the bottom of the splitter at $Z_e$ and $Q_w$. A second recycle stream from the splitter is fed back to the first anoxic reactor at $Z_r$ and $Q_r$. This model describes 8 basic biological reaction processes and considers 13 major compounds. The concentrations of these 13 compounds in the 5 reactors and 1 splitter are the state variables of the bioreactor model. The state set for the spiltter and each reactor is \{$S_{\rm I}$, $X_{\rm I}$, $S_{\rm S}$, $X_{\rm S}$, $X_{\rm B,H}$, $X_{\rm B,A}$, $X_{\rm P}$, $X_{\rm ND}$, $S_{\rm NO}$, $S_{\rm NH}$, $S_{\rm ND}$, $S_{\rm O}$, $S_{\rm ALK}$\}. The definition of these state variables can be found in \cite{Zeng2016_IECR}. There are a total of 78 states that describe the dynamics of the entire process. For the WWTP, up to 48 measurements are available for state estimation. In each reactor, eight states/state-related variables can be measured and the definitions and expressions of these measurements are shown in Table~\ref{lsy10_tabb1}. In this simulation, we detail how sensitivity analysis can be used to determine the minimum number and optimal placement of sensors, and how degree of observability can improve state estimation.

\begin{table}[!hbt] \small
 \centering
 \caption{List of measurements.}
 \label{lsy10_tabb1}
\renewcommand{\arraystretch}{1.2}
 \tabcolsep 3pt
 \begin{tabular}{ll}\hline
 Measurement                    &  Expression in the form of process states \\\hline
 Oxygen                         &  $S_{\rm O}$    \\
 Alkalinity                     &  $S_{\rm ALK}$  \\
 NH$4^+$ and NH3 nitrogen       &  $S_{\rm NH}$   \\
 Nitrate and nitrite nitrogen   &  $S_{\rm NO}$   \\
 Chemical oxygen demand (COD)   &  $S_{\rm I}$+$S_{\rm S}$+$X_{\rm I}$+$X_{\rm S}$+$X_{\rm B,H}$+$S_{\rm B,A}$ \\
 Filtered chemical oxygen demand (COD$_{\rm f}$) & $S_{\rm I}$+$S_{\rm S}$ \\
 Biological oxygen demand (BOD) &  $S_{\rm S}$+$X_{\rm S}$ \\
 Total suspended solids (TSS)   &  $X_{\rm I}$+$X_{\rm S}$+$X_{\rm B,H}$+$S_{\rm B,A}$+$X_{\rm P}$+$X_{\rm ND}$ \\\hline
\end{tabular}\end{table}

Based on the results of \cite{Zeng2016_IECR}, the initial sensor set is chosen as follows.
\begin{itemize}
\item Reactor 2: $COD_2$, $COD_{\rm f,2}$, $TSS_2$, $S_{\rm O,2}$, $S_{\rm ALK,2}$.
\item Reactor 3: $COD_3$, $COD_{\rm f,3}$, $TSS_3$, $S_{\rm O,3}$, $S_{\rm ALK,3}$.
\item Reactor 4: $S_{\rm O,4}$.
\item Reactor 5: $S_{\rm O,5}$.
\item Ideal splitter: $COD_6$, $S_{\rm NH,6}$, $S_{\rm NO,6}$, $S_{\rm ALK,6}$.
\end{itemize}

\begin{figure}[!hbt]
	\centering
	\includegraphics[width=0.8\hsize]{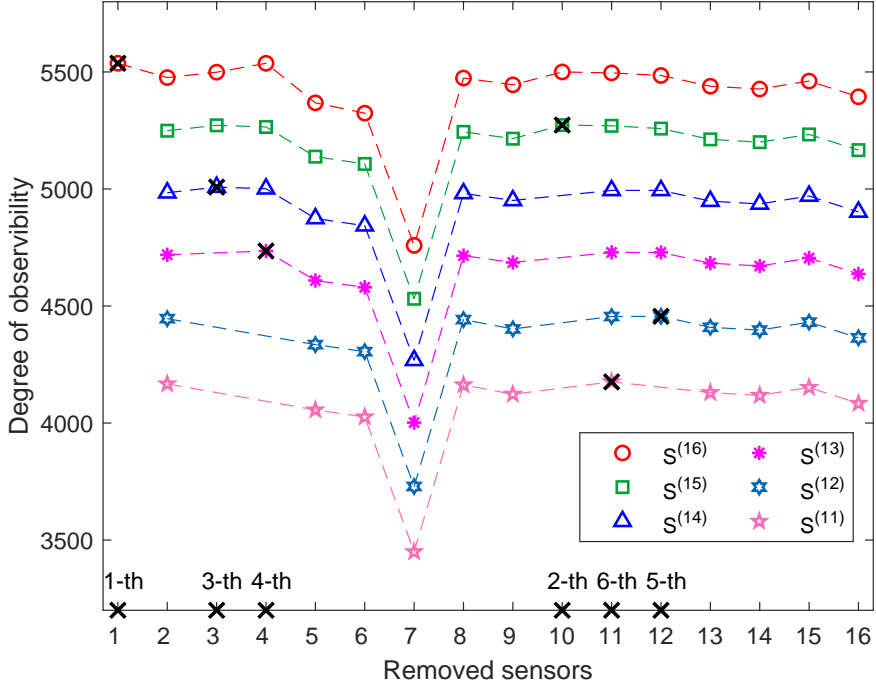}
	\caption{Degree of observability by removing different measurements from.}
	\label{lsy10_figb1}
\end{figure}

The proposed algorithm is applied to select the minimum number of sensors from the given set and to provide the optimal locations of the sensors for the minimum set. In the simulations, the process measurements are sampled every $\Delta=15$min. The data file of dry weather can be found on the International Water Association website. Initially, the sensor set $\mathcal{S}^{(16)}=\{S_{\rm O,2}$, $S_{\rm O,3}$, $S_{\rm O,4}$, $S_{\rm O,5}$, $COD_2$, $COD_3$, $COD_6$, $COD_{\rm f,2}$, $COD_{\rm f,3}$, $S_{\rm ALK,2}$, $S_{\rm ALK,3}$, $S_{\rm ALK,6}$, $TSS_2$, $TSS_3$, $S_{\rm NO,6}$, $S_{\rm NH,6}\}$ denoting as $\{1,2,...,16\}$ for simplicity, and the system is fully observable. Figure~\ref{lsy10_figb1} shows the process of removing the sensors one by one, and the corresponding degree of observability of remaining sensor combinations. It is observed that for $S^{(16)}$, when the first sensor is removed, the corresponding degree of observability of the remaining 15 sensors is the largest. This means that removing the first sensor has the smallest impact on the system observability. Therefore, sensor 1 is removed first. In the next step, another sensor is to be removed and we aim to find the optimal set containing 15 sensors. It can be seen that the tenth one is removed from the 15 sensors, and the degree of observability is the largest among all possible combinations of 14 sensors.  The remaining steps to remove sensors are performed similarly. It can be found that the 16 measurements used in \cite{Zeng2016_IECR} can be reduced to 10. If we continue to remove any of the 10 sensors, the system becomes unobservable, so the selection process is terminated at $m=10$. The selected 10 sensors are shown in Table~\ref{lsy10_tabb2} (Case 1).

\begin{figure}[!hbt]
 \centering
 \includegraphics[width=\hsize]{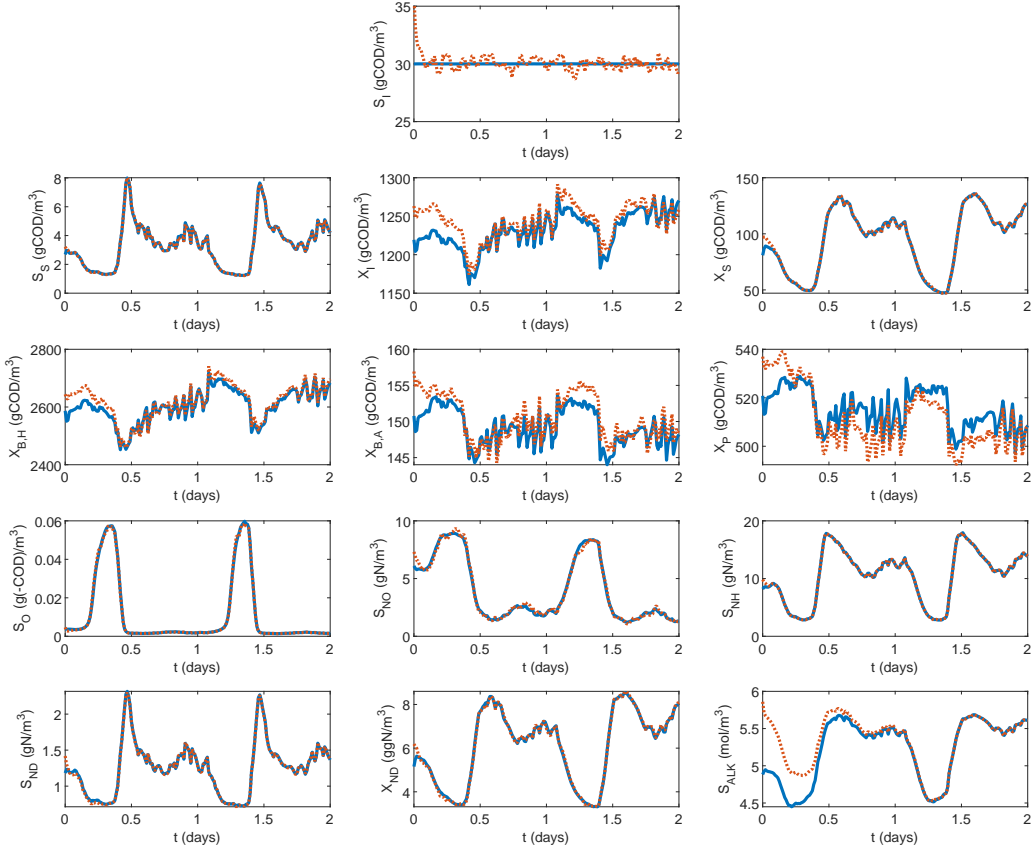}
 \caption{Trajectories of the actual process states (solid lines) and the estimates (dashed lines) with 10 measurements of reactor 2.}
\label{lsy10_figb2}
\end{figure}

Once the set of sensors is determined, we consider the state estimation using the extended Kalman filtering (EKF), the same as used in \cite{Zeng2016_IECR}, to verify the effectiveness of the selected sensors. The initial guess is set to be $1.2\bfx_s$. The tuning parameters used in the EKF are $\bfQ_w={\rm diag}((0.1\bfx_s)^2)$, $\bfR_v={\rm diag}((0.1\bfy_s)^2)$ and $\bfP(0)=\bfQ_w={\rm diag}((0.1\bfx_s)^2)$. The actual states of reactor 2 and their state estimates are shown in Figure~\ref{lsy10_figb2}. It is obvious that the state estimates can track the actual state trajectory well.

\begin{figure}[!hbt]
 \centering
 \includegraphics[width=0.8\hsize]{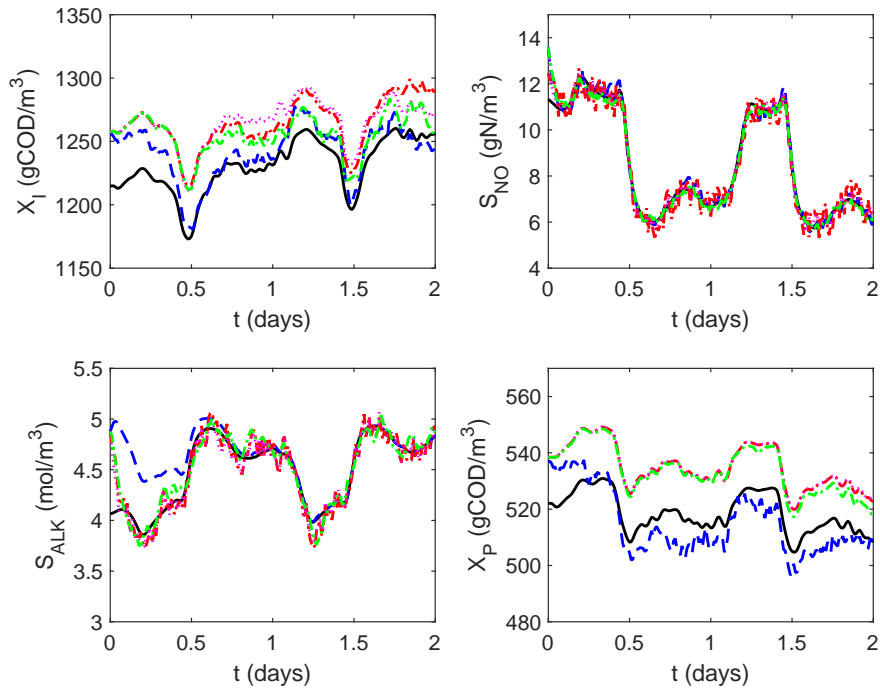}
 \caption{Trajectories of the actual process states (solid lines) and the state estimates based on 10 measurements (Case 1: blue dashed lines; Case 2: red dash-dotted lines; Case 3: dotted lines; Case 4: green dash-dotted lines) of reactor 6.}
\label{lsy10_figb3}
\end{figure}

In order to further show the validity of the selected minimum number of sensor combinations, three randomly chosen ten-sensor combinations are constructed (Cases 2-4 in Table~\ref{lsy10_tabb2}). The mean values of the normalized estimation error can be defined as
\begin{equation}
	{\rm Mean}|e|=\frac{1}{N}\sum_{t=1}^{N}e(k),
\end{equation}
where
\begin{equation}
	e(k)=\sqrt{\sum_{i=1}^{78}\left(\frac{\hat{x}_i(k)-x_i(k)}{\max(\hat{x}_i-x_i)}\right)^2}
\end{equation}
is the normalized error at sampling time $k$ based on its maximum estimation error. Mean$|e|$ and RMSE are listed in Table~\ref{lsy10_tabb2}, which are used to evaluate the performance of state estimation for each sensor combination. It can be found that the Mean$|e|$ and RMSE obtained by Case 1 are the smallest among the four cases. The trajectories of the actual states and the state estimates of reactor 6 for four cases are presented in Figure~\ref{lsy10_figb3}. From Table~\ref{lsy10_tabb2} and Figure~\ref{lsy10_figb3}, it can be found that the selected sensor set $\{2,5,6,7,8,9,13,14,15,16\}$ in Case 1 gives the best estimation results compared to other sensor combinations with the same number of sensors.

If we want to have more robust state estimation performance in the presence of significant process disturbances and measurement noise, additional sensors can be added to the minimum set of 10 sensors. In this case, it is necessary to select the sensor that contributes most to the degree of observability, and then include it in sensor set. From Figure \ref{lsy10_figb1}, we can see the order in which the sensors are removed. Therefore, the order in which the sensors are added can be based on the reverse order of removal. When we only want to add one sensor, we should select the sensor 11. If there is more budget or need to add another, then we should further select sensor 12.

\begin{table}[!hbt] \small
 \centering
 \caption{Performance comparison for four measurement configurations.}
 \label{lsy10_tabb2}
\renewcommand{\arraystretch}{1.2}
 \tabcolsep 3pt
 \begin{tabular}{lccc}\hline
        &  RMSE    & Mean$|e|$ & Sensor selection \\\hline
 Case 1 &  0.1322  & 2.1618    &  \{$S_{\rm O,3}$, $S_{\rm NH,6}$, $S_{\rm NO,6}$, $COD_2$, $COD_3$, $COD_6$, $COD_{\rm f,2}$, $COD_{\rm f,3}$, $TSS_2$, $TSS_3$\} \\
 Case 2 &  0.2066  & 3.2992    &  \{$S_{\rm O,2}$, $S_{\rm O,5}$, $S_{\rm ALK,2}$, $S_{\rm ALK,5}$, $S_{\rm NH,3}$, $S_{\rm NO,5}$, $COD_1$, $COD_{\rm f,4}$, $BOD_2$, $BOD_5$\} \\
 Case 3 &  0.2252  & 3.5492    &  \{$S_{\rm ALK,2}$, $S_{\rm ALK,3}$, $S_{\rm ALK,6}$, $S_{\rm NO,1}$, $S_{\rm NO,4}$, $COD_2$, $COD_{\rm f,1}$, $COD_{\rm f,5}$, $BOD_1$, $BOD_4$\} \\
 Case 4 &  0.1873  & 2.9660    &  \{$S_{\rm O,1}$, $S_{\rm O,4}$, $S_{\rm ALK,3}$, $S_{\rm NH,1}$, $S_{\rm NH,3}$, $S_{\rm NO,2}$, $COD_4$, $COD_{\rm f,4}$, $BOD_3$, $TSS_6$\}  \\\hline
 \end{tabular}\end{table}

\section{Conclusions}
\label{Conclusions}

In this work, a sensitivity-based method is proposed to quantify the degree of observability of nonlinear process networks, and a computationally efficient optimal sensor selection method for state estimation is developed to determine the minimum number of sensors and their optimal locations based on the degree of observability. The proposed approach is applied to a four-CSTRs process and a large-scale wastewater treatment process. The results show that the proposed algorithm is very effective in determining the minimum number of measurements and their optimal sensor locations. Compared with the existing approaches, the proposed approach does not need to solve mixed-integer programming and is very computatinally efficient. For the future work, we will consider the cost of sensors in selecting the sensors and apply the proposed method to other fields.

\section{Acknowledgement}

The first author, S.Y. Liu, is a visiting Ph.D. student in the Department of Chemical and Materials Engineering at the University of Alberta from March 2021 to February 2023. She acknowledges the financial support from the China Scholarship Council (CSC) during this period.



\begin{thebibliography}{10}

\bibitem{Shen2014_IEEETSP}
X.~Shen and P.~K. Varshney, ``Sensor selection based on generalized information
  gain for target tracking in large sensor networks,'' {\em IEEE Transactions
  on Signal Processing}, vol.~62, no.~2, pp.~363--375, 2014.

\bibitem{Anvaripour2019_TII}
M.~Anvaripour, M.~Saif, and M.~Ahmadi, ``A novel approach to reliable sensor
  selection and target tracking in sensor networks,'' {\em IEEE Transactions on
  Industrial Informatics}, vol.~16, no.~1, pp.~171--182, 2019.

\bibitem{Nahar2019_CCE}
J.~Nahar, J.~Liu, and S.~L. Shah, ``Parameter and state estimation of an
  agro-hydrological system based on system observability analysis,'' {\em
  Computers \& Chemical Engineering}, vol.~121, pp.~450--464, 2019.

\bibitem{Sahoo2019_AIChE}
S.~R. Sahoo, X.~Yin, and J.~Liu, ``Optimal sensor placement for
  agro‐hydrological systems,'' {\em AIChE Journal}, vol.~65, no.~12, 2019.

\bibitem{Qi2015_IEEETPS}
J.~Qi, K.~Sun, and W.~Kang, ``Optimal {PMU} placement for power system dynamic
  state estimation by using empirical observability gramian,'' {\em IEEE
  Transactions on Power Systems}, vol.~30, no.~4, pp.~2041--2054, 2015.

\bibitem{Abooshahab2021}
M.~A. Abooshahab, M.~Hovd, and G.~Valmorbida, ``Optimal sensor placement for
  partially known power system dynamic estimation,'' in {\em 2021 IEEE PES
  Innovative Smart Grid Technologies Europe (ISGT Europe)}, pp.~1--6, 2021.

\bibitem{Mo2011_Auto}
Y.~Mo, R.~Ambrosino, and B.~Sinopoli, ``Sensor selection strategies for state
  estimation in energy constrained wireless sensor networks,'' {\em
  Automatica}, vol.~47, no.~7, pp.~1330--1338, 2011.

\bibitem{Bajovic2011_TSP}
D.~Bajovic, B.~Sinopoli, and J.~Xavier, ``Sensor selection for event detection
  in wireless sensor networks,'' {\em IEEE Transactions on Signal Processing},
  vol.~59, no.~10, pp.~4938--4953, 2011.

\bibitem{Contreras2016_IEEETITS}
S.~Contreras, P.~Kachroo, and S.~Agarwal, ``Observability and sensor placement
  problem on highway segments: A traffic dynamics-based approach,'' {\em IEEE
  Transactions on Intelligent Transportation Systems}, vol.~17, no.~3,
  pp.~848--858, 2016.

\bibitem{Yin2018_CCE}
X.~Yin and J.~Liu, ``State estimation of wastewater treatment plants based on
  model approximation,'' {\em Computers \& Chemical Engineering}, vol.~111,
  pp.~79--91, 2018.

\bibitem{Stigter2017_SR}
J.~D. Stigter, D.~Joubert, and J.~Molenaar, ``Observability of complex systems:
  Finding the gap,'' {\em Scientific Reports}, vol.~7, no.~1, p.~16566, 2017.

\bibitem{Liu2013_PANS}
Y.-Y. Liu, J.-J. Slotine, and A.-L. Barab{\'a}si, ``Observability of complex
  systems,'' {\em Proceedings of the National Academy of Sciences}, vol.~110,
  no.~7, pp.~2460--2465, 2013.

\bibitem{Angulo2020_IEEETNSE}
M.~T. Angulo, A.~Aparicio, and C.~H. Moog, ``Structural accessibility and
  structural observability of nonlinear networked systems,'' {\em IEEE
  Transactions on Network Science and Engineering}, vol.~7, no.~3,
  pp.~1656--1666, 2019.

\bibitem{cowan2012nodal}
N.~J. Cowan, E.~J. Chastain, D.~A. Vilhena, J.~S. Freudenberg, and C.~T.
  Bergstrom, ``Nodal dynamics, not degree distributions, determine the
  structural controllability of complex networks,'' {\em PloS One}, vol.~7,
  no.~6, p.~e38398, 2012.

\bibitem{Haber2018_IEEETCNS}
A.~Haber, F.~Molnar, and A.~E. Motter, ``State observation and sensor selection
  for nonlinear networks,'' {\em IEEE Transactions on Control of Network
  Systems}, vol.~5, no.~2, pp.~694--708, 2017.

\bibitem{Haber2021_IEEETNSE}
A.~Haber, ``Joint sensor node selection and state estimation for nonlinear
  networks and systems,'' {\em IEEE Transactions on Network Science and
  Engineering}, vol.~8, no.~2, pp.~1722--1732, 2021.

\bibitem{Nugroho2021_Auto}
S.~A. Nugroho and A.~F. Taha, ``Towards understanding sensor and control nodes
  selection in nonlinear dynamic systems: Lyapunov theory meets
  branch-and-bound,'' {\em Automatica}, vol.~134, p.~109904, 2021.

\bibitem{Singh2006_IECR}
A.~K. Singh and J.~Hahn, ``Sensor location for stable nonlinear dynamic
  systems: Multiple sensor case,'' {\em Industrial \& Engineering Chemistry
  Research}, vol.~45, no.~10, pp.~3615--3623, 2006.

\bibitem{Kang2012}
W.~Kang and L.~Xu, ``Optimal placement of mobile sensors for data
  assimilations,'' {\em Tellus A: Dynamic Meteorology and Oceanography},
  vol.~64, no.~1, p.~17133, 2012.

\bibitem{Mar2003}
H.~J. Marquez, {\em Nonlinear Control Systems: Analysis and Design}, vol.~161.
\newblock John Wiley Hoboken\^{} eN. JNJ, 2003.

\bibitem{Kravaris2013_CCE}
C.~Kravaris, J.~Hahn, and Y.~Chu, ``Advances and selected recent developments
  in state and parameter estimation,'' {\em Computers \& Chemical Engineering},
  vol.~51, pp.~111--123, 2013.

\bibitem{Miao2011_SIAM}
H.~Miao, X.~Xia, A.~S. Perelson, and H.~Wu, ``On identifiability of nonlinear
  ode models and applications in viral dynamics.,'' {\em SIAM Review}, vol.~53,
  no.~1, pp.~3--39, 2011.

\bibitem{Stigter2019_Auto}
J.~D. Stigter and J.~Molenaar, ``A fast algorithm to assess local structural
  identifiability,'' {\em Automatica}, vol.~58, pp.~118--124, 2015.

\bibitem{Liu2021_IECR}
J.~Liu, A.~Gnanasekar, Y.~Zhang, S.~Bo, J.~Liu, J.~Hu, and T.~Zou,
  ``Simultaneous state and parameter estimation: the role of sensitivity
  analysis,'' {\em Industrial \& Engineering Chemistry Research}, vol.~60,
  no.~7, pp.~2971--2982, 2021.

\bibitem{LiuSY2022_ChERD}
S.~Y. Liu, X.~Yin, J.~Liu, J.~Liu, and F.~Ding, ``Distributed simultaneous
  state and parameter estimation of nonlinear systems,'' {\em Chemical
  Engineering Research and Design}, vol.~181, pp.~74--86, 2022.

\bibitem{Zhang2017_Auto}
H.~Zhang, R.~Ayoub, and S.~Sundaram, ``Sensor selection for {Kalman} filtering
  of linear dynamical systems: Complexity, limitations and greedy algorithms,''
  {\em Automatica}, vol.~78, pp.~202--210, 2017.

\bibitem{Jawaid2015_Auto}
S.~T. Jawaid and S.~L. Smith, ``Submodularity and greedy algorithms in sensor
  scheduling for linear dynamical systems,'' {\em Automatica}, vol.~61,
  pp.~282--288, 2015.

\bibitem{Qi2016}
J.~Qi, K.~Sun, and W.~Kang, ``Adaptive optimal {PMU} placement based on
  empirical observability gramian,'' {\em IFAC-PapersOnLine}, vol.~49, no.~18,
  pp.~482--487, 2016.

\bibitem{Grubben2018_IJC}
N.~L. Grubben and K.~J. Keesman, ``Controllability and observability of 2{D}
  thermal flow in bulk storage facilities using sensitivity fields,'' {\em
  International Journal of Control}, vol.~91, no.~7, pp.~1554--1566, 2018.

\bibitem{Rashedi2018_TII}
M.~Rashedi, J.~Liu, and B.~Huang, ``Triggered communication in distributed
  adaptive high-gain {EKF},'' {\em IEEE Transactions on Industrial
  Informatics}, vol.~14, no.~1, pp.~58--68, 2017.

\bibitem{Alex2008}
J.~Alex, L.~Benedetti, J.~Copp, K.~Gernaey, U.~Jeppsson, I.~Nopens, M.-N. Pons,
  J.-P. Steyer, and P.~Vanrolleghem, {\em Benchmark Simulation Model no. 1
  (BSM1)}, vol.~TEIE-7229.
\newblock Lund University, 2008.

\bibitem{Zeng2016_IECR}
J.~Zeng, J.~Liu, T.~Zou, and D.~Yuan, ``Distributed extended {Kalman} filtering
  for wastewater treatment processes,'' {\em Industrial \& Engineering
  Chemistry Research}, vol.~55, no.~28, pp.~7720--7729, 2016.

\end{thebibliography}

\end{document}